\documentclass[english]{paper}
\usepackage[T1]{fontenc}
\usepackage[latin9]{inputenc}
\usepackage[letterpaper]{geometry}
\usepackage{amsthm}
\usepackage{graphicx}

\newcommand{\lyxdot}{.}

\theoremstyle{plain}
\newcommand{\lyxaddress}[1]{
\par {\raggedright #1
\vspace{1.4em}
\noindent\par}
}

\usepackage{babel}

\begin{document}

\title{Generalization learning in a perceptron with binary synapses}

\author{Carlo Baldassi}

\maketitle

\lyxaddress{Politecnico di Torino\\
Dipartimento di Fisica\\
Corso Duca degli Abruzzi 24\\
I-10129, Torino, Italy\\
email: carlo.baldassi@polito.it}
\begin{abstract}
We consider the generalization problem for a perceptron with binary
synapses, implementing the Stochastic Belief-Propagation-Inspired
(SBPI) learning algorithm which we proposed earlier, and perform a
mean-field calculation to obtain a differential equation which describes
the behaviour of the device in the limit of a large number of synapses
$N$. We show that the solving time of SBPI is of order $N\sqrt{\log N}$,
while the similar, well-known clipped perceptron (CP) algorithm does
not converge to a solution at all in the time frame we considered.
The analysis gives some insight into the ongoing process and shows
that, in this context, the SBPI algorithm is equivalent to a new,
simpler algorithm, which only differs from the CP algorithm by the
addition of a stochastic, unsupervised meta-plastic reinforcement
process, whose rate of application must be less than $\sqrt{2/\left(\pi N\right)}$
for the learning to be achieved effectively. The analytical results
are confirmed by simulations. 
\emph{subclass:} 68T05, 68T15, 82C32\end{abstract}

\section{Introduction}

\label{intro} The perceptron was first proposed by Rosenblatt\cite{rosenblatt62}
as an extremely simplified model of a neuron, consisting in a number
$N$ of input lines, each one endowed with a weight coefficient (representing
the individual synaptic conductances), all of which converge in a
central unit (representing the soma) with a single output line (the
axon). Typically, the output is computed from a threshold function
of the weighted sum of the inputs, and the time in the model is discretized,
so that, at each time step, the output does only depend on the input
at that time step. The unit can adapt its behaviour over time by modifying
the synaptic weights (and possibly the output threshold), and thus
it can undergo a learning (or memorizing) process. In this paper,
we only consider the case in which the learning process is {}``supervised'',
i.e.~in which there is some feedback from the outside, typically
in the form of an {}``error signal'', telling the unit that the
output is wrong, as opposed to {}``unsupervised'' learning, in which
no feedback is provided to the unit.

Despite its simplicity, the perceptron model is very powerful, being
able to process its inputs in parallel, to retain an extensive amount
of information, and to plastically adapt its output over time in an
on-line fashion. Furthermore, it displays a highly non-trivial behaviour,
so that much research has been devoted to the study of its analytical
properties and of the optimal learning strategies in different contexts
(see e.g.~\cite{baldassi07efficient,engel01,fusi05,gutfreund90b,kinouchi92optimalgeneralization,krauth89,Solla98optimalperceptron}
and references therein). In the supervised case, there are typically
two scenarios: the first, which we will call {}``classification''
problem in the rest of this paper, is defined by a given set of input-output
associations that the unit must learn to reproduce without errors,
while the second, which we will call {}``generalization'' problem,
is defined by a given input-output rule that the unit must learn to
implement as closely as possible. Here, we will mainly focus our attention
on this last problem.

Furthermore, we will restrict to the case in which the synaptic weights
are assumed to be binary variables. Binary models are inherently simpler
to implement and more robust over time against noise with respect
to models in which the synaptic weights are allowed to vary over a
continuous set of values, while having a comparable information storage
capacity\cite{engel01,gutfreund90b,krauth89}; furthermore, some recent
experimental results\cite{oconnor05,petersen98}, as well as some
arguments from theoretical studies and computer simulations\cite{bhalla99,bialek2000stabilitynoise,hayer05molecularswitches,miller05},
suggest that binary-synapses models could also be more relevant than
continuous ones as neuronal models exhibiting long term plasticity.
However, from an algorithmic point of view, learning is much harder
in models with binary synapses than in in models with continuous synapses:
in the worst-case scenario, the classification learning problem is
known to be NP-complete for binary weights\cite{blum92}, while it
is easy to solve it effectively with continuous weights\cite{engel01}.
Even in the case of random, uncorrelated inputs-output associations,
the solution space of the classification problem in the binary case
is in a broken-symmetry phase, while in the continuous case it is
not\cite{krauth89}, implying that the learning strategies which successfully
solve the learning problem in the latter case are normally not effective
in the former.

Despite these difficulties, an efficient, easily implementable, on-line
learning algorithm can be devised which solves efficiently the binary
classification problem in the case of random, uncorrelated input stimuli\cite{baldassi07efficient}.
Such an algorithm was originally derived from the standard Belief
Propagation algorithm\cite{braunstein06,kabashima04,yedidia03}, and
hence named `Stochastic Belief Propagation-Inspired' (SBPI). The SBPI
algorithm makes an additional requirement on the model, namely, that
each synapse in the device, besides the weight, has an additional
hidden, discretized internal state; transitions between internal states
may be purely meta-plastic, meaning that the synaptic strength does
not necessarily change in the process, but rather that the plasticity
of the synapse does. The SBPI learning rules and hidden states requirements
are the same as those of the well known clipped perceptron algorithm
(CP, see e.g.~\cite{rosenzvi00}), the only difference being an additional,
purely meta-plastic rule, which is only applied if the answer given
by the device is correct, but such that a single variable flip would
result in a classification error.

The SBPI algorithm was derived and tested in the context of the classification
problem: in such scheme, all the input patterns are extracted from
a given pattern set, randomly generated before the learning session
takes place, and presented to the student device repeatedly, the outcome
being compared to the desired one, until no classification errors
are made any more. Since the analytical treatment of the learning
process is awkward in such case, due to the temporal correlations
emerging in the input patterns as a consequence of the repeated presentations,
we could only test the SBPI algorithm performance by simulations,
and compare it to that of other similar algorithms, such as the CP
algorithm and the cascade model \cite{fusi05}. It turned out that
the additional learning rule which distiguishes the SBPI algorithm
from the CP algorithm is essential to SBPI's good performance, and
that there exists an optimal number of parameters for both the number
of internal hidden states per synapse and for the rate of application
of the novel learning rule.

In order to understand the reason for the SBPI new rule's effectiveness,
it is necessary to give an analytical description of the learning
process under the CP and SBPI learning rules; to this end, we consider
here the problem of generalization from examples: in such scheme,
the input patterns are generated afresh at each time step, and the
goal is to learn a linearly separable classification rule, provided
by a teacher device. Perfect learning is achieved if the student's
synapses match the teacher's ones. Using a teacher unit identical
to the student to define the goal input-output function ensures that
a solution to the problem exists and is unique, but we also briefly
consider the case of a non-perfect-learnable rule, provided by a continuous-weights
teacher perceptron. The learning process in this case is much easier
to treat analytically than in the classification case, since the input
patterns are not temporally correlated, and the dynamical equations
for this system can be derived by a mean field calculation for the
case of a learnable rule. This problem is in fact easier to address
than the classification problem, and optimal algorithms can be found
which solve it in the binary synapses case as well (see e.g.~\cite{golea90,rosenzvi00,Solla98optimalperceptron});
however, such algorithms are not suitable to be considered as candidates
for biological models for online learning, being either too complex
or requiring to perform all intermediate operations with an auxiliary
device with continuous synaptic weights.

The resulting differential equation set that we obtained gives some
insight on the learning dynamics and about the reason for SBPI's effectiveness,
and allows for a further simplification of the SBPI algorithm, yielding
an even more attractive model of neuronal unit, both from the point
of view of biological feasibility and of hardware manufacturing design
simplicity. With a special choice for the parameters, the solution
to the equation set is simple enough to be studied analytically and
demonstrate that the algorithm converges in a number of time steps
which goes as $N\sqrt{\log N}$. All the results are confirmed by
simulations.

The outline of the rest of this paper is as follows: in Sects.~2
and 3 we define in detail the learning algorithm and the generalization
problem, respectively. In Sects.~4 and 5 we derive the mean-field
dynamics for the CP and SBPI algorithms, and in Sect.~6 we derive
the set of continuous differential equations which describes the process
in the $N\to\infty$ limit and exhibit a solution. In Sect.~7 we
consider a special case in which the equation set can be simplified
and derive some analytical results on convergence time in such case.
In Sect.~8 we consider the case of bounded hidden states. In Sect.~9
we briefly consider the case of a non learnable rule. In Sect.~10
we discuss the simplified algorithm derived in Sect.~\ref{sec:Histogram-dynamics-for-the-SBPI-algorithm}.
We summarize our results in the last section.

\section{The SBPI learning algorithm}

\label{sec:The-SBPI-learning-algorithm} The device we consider is
a binary perceptron with $N$ synapses, each of which can take the
values $w_{i}=\pm1$, receiving inputs $\xi_{i}^{\mu}=\pm1$, with
output $\sigma^{\mu}=\pm1$, and threshold $\theta=0$. Thus, the
device output is given as a function of the inputs and of the internal
state as\[
\sigma^{\mu}=\mbox{sign}\left(\sum_{i=1}^{N}w_{i}\xi_{i}^{\mu}\right)\]

Furthermore, each synapse is endowed with a discretized internal variable
$h_{i}$, which only plays an active role during the learning process;
for simplicity, we will consider it to be an odd-valued integer. At
any given time, the sign of this quantity gives the value of the corresponding
synaptic weight, $w_{i}=\mbox{sign}\left(h_{i}\right)$. We will start
by considering the case of unbounded hidden states, and then turn
to the bounded case.

SBPI is an on-line supervised learning algorithm; upon presentation
of a pattern $\left\{ \xi_{i}^{\mu},\sigma_{D}^{\mu}\right\} $, where
$\sigma_{D}^{\mu}$ is the desired output, the stability is computed
as\[
\Delta^{\mu}=\sigma_{D}^{\mu}\left(\sum_{i=1}^{N}w_{i}\xi_{i}^{\mu}\right)\]

The way synaptic weights are updated depends on the value of $\Delta^{\mu}$:
\begin{enumerate}
\item If $\Delta^{\mu}>\theta_{m}$, then nothing is done 
\item If $0<\Delta^{\mu}\le\theta_{m}$, then only the synapses for which
$w_{i}=\xi_{i}^{\mu}\sigma_{D}^{\mu}$, are updated, and only with
probability $p_{s}$. 
\item If $\Delta^{\mu}\le0$, then all synapses are updated 
\end{enumerate}
Here, $\theta_{m}$ is a secondary threshold, expressed as an even
integer, and $p_{s}\in\left[0,1\right]$. The update rule applies
to the hidden synaptic variables: \[
h_{i}\to h_{i}+2\xi_{i}^{\mu}\sigma_{D}^{\mu}\]

The factor $2$ is required in order to keep the value of the hidden
variables odd, which in turn is useful for avoiding the ambiguous,
but otherwise immaterial, $h_{i}=0$ case. Note that the only actual
plasticity events occur when the hidden variables change sign; also,
the update in rule 2 is always in the direction of increasing the
hidden variables' modulus, thus reinforcing the synaptic value by
making it less likely to switch.

When the probability $p_{s}$ or, equivalently, when the secondary
threshold $\theta_{m}$ are set to $0$, rule 2 is never applied and
the algorithm is reduced to the CP algorithm.

In the special case, $p_{s}=1$ and $\theta_{m}=2$, we refer to the
algorithm as to BPI.

\section{Definition of the generalization learning problem}

\label{sec:Definition-of-the-generalization-learning-problem} The
protocol which was originally used to obtain the SBPI update rules
was that of classification of random patterns extracted from a given
set; learning of the correct classification was achieved by repeated
presentations of the patterns from the set and application of the
update rules. The maximum number of input-output associations that
the system could memorize in this way was shown by simulations to
be proportional to the number of synapses $N$, the coefficient of
proportionality being fairly close to the maximal theoretical value,
with an order $O\left(\log\left(N\right)^{1.5}\right)$ presentations
per pattern required on average.

Here instead we will consider the problem of learning a rule from
a teacher perceptron, identical to the student (the case of a different
teacher device being considered in Sec.~\ref{sec:Teacher-with-continuous-synaptic-weights});
the patterns are generated at random at each time step, each input
$\xi_{i}$being extracted independently with probability $P\left(\xi_{i}=+1\right)=P\left(\xi_{i}=-1\right)=1/2$,
and the desired output is given by the teacher. Thus, the goal is
to reach a perfect overlap with the teacher, an event which can be
thought of as the student having learned an association rule. An optimal
learning algorithm for this problem, which reaches the solution in
about $1.245N$ steps in the limit of large $N$, can be derived by
the Bayesian approach\cite{Solla98optimalperceptron} (which is equivalent
to the Belief Propagation approach \cite{braunstein06} in this case);
however, this optimal algorithm does not work in an on-line fashion,
as it requires to keep the memory of each pattern which was presented
thus far to the device. An on-line approximation of the optimal algorithm,
proposed in \cite{Solla98optimalperceptron} and later re-derived
from a different approach in \cite{baldassi07efficient} as an intermediate
step towards SBPI, overcomes this problem at the expense of a lower
performance, but it still requires the internal storage of continuous
quantities, and complex computations to be performed at each time
step.

In order to simplify the notation in the rest of this paper, we will
assume that the student is always trained only on patterns whose desired
output is $+1$, which can be insured in this way: at each time $\tau$
a new pattern $\left\{ \chi_{i}^{\tau}\right\} _{i}$ is generated
randomly and presented to the teacher, whose output is $\sigma_{T}^{\tau}$;
then, the pattern $\left\{ \xi_{i}^{\tau}\right\} =\left\{ \sigma_{T}^{\tau}\chi_{i}^{\tau}\right\} $
is presented to the student, with desired output $\sigma_{D}^{\tau}=+1$.
Also, we can assume, without loss of generality, that all the teacher's
synapses are set to $w_{i}^{T}=+1$. This implies that the student
will only be presented patterns in which there are more positive than
negative inputs.

In the following, we shall show that it is possible to describe the
average learning dynamics and estimate the time needed for the student
to reach overlap $1$ with the teacher, $q=\frac{1}{N}\left(w\cdot w^{T}\right)=1$.

\section{Histogram dynamics for the CP algorithm}

\label{sec:Histogram-dynamics-for-the-CP-algorithm} We will do a
mean-field-like approximation to the problem: at each time step, given
the histogram of the hidden variables at a time $\tau$, $\mbox{P}^{\tau}\left(\left\{ h_{i}\right\} \right)$,
we compute the average distribution (over the input patterns) at time
$\tau+1$, $P^{\tau+1}\left(\left\{ h_{i}\right\} \right)$, and iterate.
The approximation here resides in the fact that, at each step, what
we obtain is an average quantity (a single histogram), which we use
as input in the following step, while a more complete description
would involve the evolution of the whole probability distibution over
all the possible resulting histograms. Therefore, we are implicitly
assuming that the spread of such probability distribution around its
average is negligible; our results confirm this assumption.

We will start from the simpler case of the CP algorithm (no rule 2),
and temporarily drop the index $\tau$.

Let us first compute the probability of making a classification error.
This only depends on the current teacher-student overlap $q$. We
will denote by $q_{+}$ ($q_{-}$) the fraction of student synapses
which are set to $+1$ ($-1$), so that the overlap is $q=q_{+}-q_{-}=2q_{+}-1$.
In the following, we have to consider separately the $+1$ and $-1$
synapses: we denote by $\nu_{+}$ the number of positive inputs over
the positive synapses, and by $\nu_{-}$ the number of positive inputs
over the negative synapses. Because of the constraint on the patterns,
there have to be more positive inputs than negative ones, i.e.~$\nu_{+}+\nu_{-}>\frac{N}{2}$.
The perceptron will classify the pattern correctly if $\nu_{+}+\left(q_{-}N-\nu_{-}\right)>\frac{N}{2}$,
thus the probability that the student makes an error is given by\begin{eqnarray*}
p_{e} & = & 2\int d\mu\left(\nu_{+}\right)d\mu\left(\nu_{-}\right)\Theta\left(\nu_{+}+\nu_{-}-\frac{N}{2}\right)\cdot\\
 & \cdot & \Theta\left(-\left(\nu_{+}+\left(q_{-}N-\nu_{-}\right)-\frac{N}{2}\right)\right)\end{eqnarray*}
 where $\mu\left(\nu_{\pm}\right)$ is the measure over $\nu_{\pm}$
without the constraint on the pattern (which is explicitly obtained
by cutting half of the cases and renormalizing). In the large $N$
limit, this is a normal distribution, centered on $\frac{q_{\pm}N}{2}$
with variance $\frac{q_{\pm}N}{4}$, thus we can write the above probability
as\begin{eqnarray}
p_{e} & = & 2\int Dx_{+}Dx_{-}\,\Theta\left(\sqrt{q_{+}}x_{+}+\sqrt{q_{-}}x_{-}\right)\Theta\left(-\sqrt{q_{+}}x_{+}+\sqrt{q_{-}}x_{-}\right)\nonumber \\
 & = & 1-\frac{2}{\pi}\arctan\left(\sqrt{\frac{q_{+}}{q_{-}}}\right)\nonumber \\
 & = & \frac{1}{\pi}\arccos\left(q\right)\label{eq:perr}\end{eqnarray}
 where we used the shorthand notation $Dx=dx\frac{1}{\sqrt{2\pi}}e^{-\frac{x^{2}}{2}}$
(eq.~\ref{eq:perr} is the standard relation between the generalization
error and the teacher-student overlap in perceptrons, see e.g.~\cite{engel01}).

We then focus on a synapse with negative value, and compute the probability
that there is an error and that the synapse receives a positive input:\begin{eqnarray*}
 &  & P\left(\Delta<0\wedge\xi_{i}=1|w_{i}=-1\right)=\\
 & = & 2\int Dx_{+}Dx_{-}\,\left(\frac{1}{2}+\frac{x_{-}}{2\sqrt{q_{-}N}}\right)\Theta\left(\sqrt{q_{+}}x_{+}+\sqrt{q_{-}}x_{-}\right)\Theta\left(-\sqrt{q_{+}}x_{+}+\sqrt{q_{-}}x_{-}\right)\\
 & = & \frac{p_{e}}{2}+\frac{1}{\sqrt{2\pi N}}+\mathcal{O}\left(\frac{1}{N}\right)\end{eqnarray*}

The probability that a negative-valued synapse receives a negative
input, and that an error is made, is very similar:\[
P\left(\Delta<0\wedge\xi_{i}=-1|w_{i}=-1\right)=\frac{p_{e}}{2}-\frac{1}{\sqrt{2\pi N}}+\mathcal{O}\left(\frac{1}{N}\right)\]

The probabilities for positive-valued synapses are simpler:\[
P\left(\Delta<0\wedge\xi_{i}=\pm1|w_{i}=+1\right)=\frac{p_{e}}{2}+\mathcal{O}\left(\frac{1}{N}\right)\]

Therefore, a positive-valued synapse (which is thus correctly set
with respect to the teacher) has an equal probability of switching
up or down one level, while a negative-valued one (which is thus wrongly
set) has a higher probability of switching up than down. The histogram
dynamics can be written in a first-order approximation as:\begin{eqnarray}
P^{\tau+1}\left(h\right) & = & P^{\tau}\left(h\right)\left[1-p_{e}^{\tau}\right]\nonumber \\
 &  & +\, P^{\tau}\left(h+2\right)\left[\frac{p_{e}^{\tau}}{2}-\frac{\Theta\left(-\left(h+2\right)\right)}{\sqrt{2\pi N}}\right]\nonumber \\
 &  & +\, P^{\tau}\left(h-2\right)\left[\frac{p_{e}^{\tau}}{2}+\frac{\Theta\left(-\left(h-2\right)\right)}{\sqrt{2\pi N}}\right]\label{eq:histdynCP}\end{eqnarray}

where, as usual, the $h$'s are assumed do be odd. It can be easily
verified that normalization is preserved by this equation.

Note that, if $p_{e}$ is very small, $\frac{p_{e}}{2}-\frac{1}{\sqrt{2\pi N}}$
may become negative, which is meaningless; in terms of the overlap,
this happens when $q_{-}N<\frac{\pi}{2}$, i.e.~when convergence
is reached up to just one or two synapses (in fact, this does never
happen with the CP algorithm, which does not appear to ever converge
to the solution or to even get close to convergence). This is due
to the fact that the gaussian approximation we used is not valid any
longer when $q_{-}$ is of order $N^{-1}$; note however that this
is not really an issue for practical purposes, as simulations show
that whenever the algorithm gets into this region, convergence is
eventually reached in short time.

\section{Histogram dynamics for the SBPI algorithm}

\label{sec:Histogram-dynamics-for-the-SBPI-algorithm} We now turn
to SBPI. We have to compute the probability that the new rule 2 is
applied, which happens when $0<\Delta\le\theta_{m}$ with probability
$p_{s}$; thus:\begin{eqnarray}
p_{b} & = & 2p_{s}\int Dx_{+}Dx_{-}\,\Theta\left(\sqrt{q_{+}}x_{+}+\sqrt{q_{-}}x_{-}\right)\Theta\left(\sqrt{q_{+}}x_{+}-\sqrt{q_{-}}x_{-}\right)\nonumber \\
 &  & \times\,\Theta\left(-\sqrt{q_{+}}x_{+}+\sqrt{q_{-}}x_{-}+\frac{\theta_{m}}{\sqrt{N}}\right)\nonumber \\
 & = & \frac{p_{s}\theta_{m}}{\sqrt{2\pi N}}+\mathcal{O}\left(\frac{1}{N}\right)\label{eq:pb}\end{eqnarray}

The leading term is of order $N^{-\frac{1}{2}}$, so there's no need
to distinguish between positive and negative synapses here, because
the difference between the two cases is of order $N^{-1}$. Thus,
each synapse has a probability $p_{b}/2$ of moving away from $0$
and a probability $p_{b}/2$ of standing still, since only half of
the synapses are involved in rule 2 each time it is applied.

We note that the result does not depend on the internal state of the
device: it is a constant, acting for both positive and negative synapses.
Furthermore, we see that we can reduce the number of parameters by
defining \begin{equation}
k=p_{s}\theta_{m}\label{eq:k}\end{equation}

This means that, in the generalization context and in the limit of
large $N$, rule 2 in the SBPI algorithm can be substituted by a stochastic,
generalized and unsupervised reinforcement process. We shall come
back to this issue in Sec.~\ref{sec:Simplified-algorithm}.

Using eq.~\ref{eq:pb} we can add rule 2 to eq.~\ref{eq:histdynCP},
getting the full SBPI dynamics:\begin{eqnarray}
P^{\tau+1}\left(h\right) & = & P^{\tau}\left(h\right)J_{0}^{\tau}+P^{\tau}\left(h+2\right)J_{-}^{\tau}\left(h+2\right)+P^{\tau}\left(h-2\right)J_{+}^{\tau}\left(h-2\right)\label{eq:histdynSBPI}\end{eqnarray}
where\begin{eqnarray}
J_{0}^{\tau} & = & 1-p_{e}^{\tau}-\frac{k/2}{\sqrt{2\pi N}}+\nonumber \\
J_{-}^{\tau}\left(h\right) & = & \frac{p_{e}^{\tau}}{2}-\Theta\left(-h\right)\frac{1}{\sqrt{2\pi N}}+\Theta\left(h\right)\frac{k/2}{\sqrt{2\pi N}}\label{eq:histdynSBPIdeltas}\\
J_{+}^{\tau}\left(h\right) & = & \frac{p_{e}^{\tau}}{2}+\Theta\left(-h\right)\frac{1}{\sqrt{2\pi N}}+\Theta\left(h\right)\frac{k/2}{\sqrt{2\pi N}}\nonumber \end{eqnarray}

The agreement between this formula and the simulations is almost perfect,
except when the average number of wrong synapses is very low, i.e.~when
$q_{-}N$ is of order $1$, as can be seen in Fig.~\ref{fig:comparisons}.

\section{Continuous limit}

\label{sec:Continuous-limit} Equations \ref{eq:histdynSBPI} and
\ref{eq:histdynSBPIdeltas} can be converted to a continuous equation
in the large $N$ limit, by rescaling the variables:\begin{eqnarray}
t & = & \frac{\tau}{N}\\
x & = & \frac{h}{\sqrt{N}}\end{eqnarray}
 and using a probability density\begin{eqnarray}
p\left(x,t\right) & = & \sqrt{N}P^{Nt}\left(\sqrt{N}x\right)\end{eqnarray}

Note that the $\sqrt{N}$ scaling of the hidden variables is the same
which we found empirically in the classification learning problem\cite{baldassi07efficient}.

Using these and taking the limit $N\to\infty$ we get the partial
differential equation set:\begin{eqnarray}
\frac{\partial p}{\partial t}\left(x,t\right) & = & 2p_{e}\left(t\right)\frac{\partial^{2}p}{\partial x^{2}}\left(x,t\right)\nonumber \\
 &  & -\,\frac{1}{\sqrt{2\pi}}\frac{\partial p}{\partial x}\left(x,t\right)\left[\left(4-k\right)\Theta\left(-x\right)+k\Theta\left(x\right)\right]+\nonumber \\
 &  & +\,\delta\left(x\right)\Theta\left(-x\right)\gamma^{-}\left(t\right)+\delta\left(x\right)\Theta\left(x\right)\gamma^{+}\left(t\right)\label{eq:pde}\\
p_{e}\left(t\right) & = & \frac{1}{\pi}\arccos\left(q\left(t\right)\right)\\
q\left(t\right) & = & 2\int_{0}^{\infty}dx\, p\left(x,t\right)-1\end{eqnarray}
 where $\delta\left(x\right)$ represents the Dirac delta function.

The two quantities $\gamma^{-}\left(t\right)$ and $\gamma^{+}\left(t\right)$
don't need to be written explicitly, since they can be specified by
imposing two conditions on the solution, normalization and continuity:\begin{eqnarray}
\int_{-\infty}^{+\infty}p\left(x,t\right) & = & 1\label{eq:norm}\\
p\left(0^{-},t\right) & = & p\left(0^{+},t\right)\label{eq:cont}\end{eqnarray}

The reason for the continuity requirement, eq.~(\ref{eq:cont}),
is the following: if there would be a discontinuity in $x=0$, the
net probability flux through that point would diverge, as can be seen
by direct inspection of eq.~(\ref{eq:histdynSBPI}) and considering
how the $\tau$ and $h$ variables scale with $N$. Note that, in
the BPI case $k=2$, enforcing these two constraints simply amounts
at setting $\gamma^{\pm}\left(t\right)=0$, as discussed in the next
section.

As a whole, eq.~(\ref{eq:pde}) is non-local, since the evolution
in each point depends on what happens at $x=0$; on the other hand,
it greatly simplifies away from that point: on either side of the
$x$ axis, it reduces to a Fokker-Planck equation, with a time-dependent
coefficient of diffusion, and a constant drift term. In general, the
drift term is different between the left and right side of the $x$
axis, and depends on $k$; this difference gives rise to an accumulation
of the probability distribution on both sides of the point $x=0$
(expressed by the two Dirac delta functions in the equation).

For negative $x$, equation \ref{eq:pde} reads:\begin{eqnarray}
\frac{\partial p}{\partial t}\left(x,t\right) & = & 2p_{e}\left(t\right)\frac{\partial^{2}p}{\partial x^{2}}\left(x,t\right)-\frac{\left(4-k\right)}{\sqrt{2\pi}}\frac{\partial p}{\partial x}\left(x,t\right)\label{eq:left}\end{eqnarray}

If the initial distribution, at time $t_{0}$, is a gaussian centered
in $x_{0}$ with variance $v_{0}$, then the solution to this equation
is a gaussian whose center $\bar{x}\left(t\right)$ and variance $v\left(t\right)$
obey the equations:\begin{eqnarray}
\bar{x}\left(t\right) & = & x_{0}+\frac{4-k}{\sqrt{2\pi}}\left(t-t_{0}\right)\\
v\left(t\right) & = & v_{0}+4\int_{t_{0}}^{t}dt^{\prime}p_{e}\left(t^{\prime}\right)\end{eqnarray}

Let us call $g^{-}\left(x,t,t_{0}\right)$ such a solution, assuming
$x_{0}=0$ and $v_{0}=0$ (i.e.~assuming the initial state to be
a Dirac-delta centered in $0$). We can define in an analogous way
a solution to the $x>0$ branch of equation \ref{eq:pde}:

\begin{eqnarray}
\frac{\partial p}{\partial t}\left(x,t\right) & = & 2p_{e}\left(t\right)\frac{\partial^{2}p}{\partial x^{2}}\left(x,t\right)-\frac{k}{\sqrt{2\pi}}\frac{\partial p}{\partial x}\left(x,t\right)\label{eq:right}\end{eqnarray}

As before, this equation transforms gaussians into gaussians: the
corresponding solution $g^{+}\left(x,t,t_{0}\right)$ only differs
from $g^{-}$ in that the centre of the gaussian moves to the right
with a velocity proportional to $k$, rather than $4-k$.

\begin{figure}
\includegraphics[width=0.75\columnwidth]{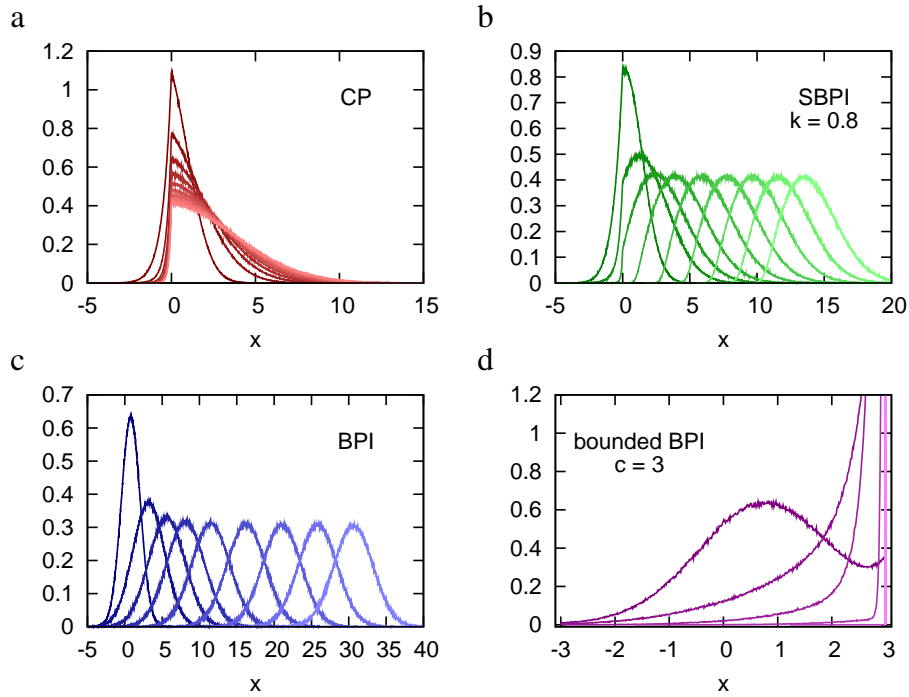} 

\caption{Evolution of the histograms with time (dark lines to light lines,
taken in time steps of $\Delta t=3$, from $t=1$ to $t=25$), in
simulations with four different algorithms ($500$ samples at $N=32001$).
In panels \textbf{a} and \textbf{b}, the positive and negative sides
of the curve obey different differential equations; in the CP algorithm
there's no drift term on the right side, and thus the majority of
the synapses stays near zero, causing a significant fraction of the
synapses to be pushed back to the negative side. The distributions
are gaussians for the unbounded BPI algorithm (panel \textbf{c}),
while setting a boundary makes the histograms accumulate at the boundary
(panel \textbf{d}). In all cases, the initial distribution was random,
with all the synapses at $h=\pm1$.}

\label{fig:hist-evol} 
\end{figure}

Overall, this gives a qualitative understanding of what happens during
learning: away form $x=0$, on both sides there's a diffusion term
(the same for both), which tends to $0$ if the majority of the synapses
gets to the right side of the $x$ axis. The synapses are `pushed'
right by the drift with `strength' $k$ on the right side and $4-k$
on the left side. Right at $x=0,$ there's a bi-directional flux between
the two sides of the solution, such that the overall area is conserved
and that the curve is continuous (even if the derivatives are not).
Thus, it is evident that both $k\le0$ and $k\ge4$ are very poor
choices (and they include the CP algorithm, which corresponds to $k=0$).
If the majority of the synapses eventually reaches the right side,
the diffusion stops and the drift dominates. The evolution of the
histograms at different times for different values of $k$ is shown
in Fig.~\ref{fig:hist-evol}.

An analytical solution to eq.~(\ref{eq:pde}) can be written in terms
of the functions $g^{\pm}$ defined above: the flux through $x=0$
gives rise, in the continuous limit, to the generation of Dirac deltas
in the origin, which in turn behave like gaussians of $0$ variance
that start to spread and shift. Due to the homogeneity of the equation,
this allows to write a solution as a weighted temporal convolution
of evolving gaussians: first, we write the initial condition as $p\left(x,0\right)=p_{0}\left(x\right)$;
then, we define $p_{0}^{-}\left(x,t\right)$ as the time evolution
of $p_{0}\left(x\right)$ under eq.~(\ref{eq:left}) and $p_{0}^{+}\left(x,t\right)$
as the time evolution of $p_{0}\left(x\right)$ under eq.~(\ref{eq:right})
(these can normally be computed easily, e.g.~by means of Fourier
transforms). This allows us to write the solution in the form:\begin{equation}
p\left(x,t\right)=\Theta\left(-x\right)p^{-}\left(x,t\right)+\Theta\left(x\right)p^{+}\left(x,t\right)\end{equation}
 where \[
p^{\pm}\left(x,t\right)=p_{0}^{\pm}\left(x,t\right)+\int_{0}^{t}dt^{\prime}\,\gamma^{\pm}\left(t^{\prime}\right)g^{\pm}\left(x,t,t^{\prime}\right)\]
 with the constraints given in eqs.~(\ref{eq:norm}) and (\ref{eq:cont}).
This solution can be verified by direct substitution in eq.~(\ref{eq:pde});
it is not likely to be amenable to further analytical treatment, but
it is sufficient for numerical integration, which indeed shows an
almost perfect agreement with the data obtained through histogram
evolution at large $N$, as shown in Fig.~\ref{fig:comparisons}a.

\section{Density evolution for BPI}

\label{sec:Density-evolution-for-BPI} In the BPI case, i.e.~when
$k=2$, the two sides of eq.~(\ref{eq:pde}) are equal; thus, the
terms $\gamma^{\pm}\left(t\right)$ both vanish, and eq.~(\ref{eq:pde})
simplifies to:

\begin{eqnarray}
\frac{\partial p}{\partial t}\left(x,t\right) & = & 2p_{e}\left(t\right)\frac{\partial^{2}p}{\partial x^{2}}\left(x,t\right)-\sqrt{\frac{2}{\pi}}\frac{\partial p}{\partial x}\left(x,t\right)\end{eqnarray}

If the initial distribution is a gaussian centered in $x_{0}$ with
variance $v_{0}$, $p\left(x,0\right)=\frac{1}{\sqrt{v_{0}}}G\left(\frac{x-x_{0}}{\sqrt{v_{0}}}\right)$,
then the evolution of the distribution is described by the following
system of equations:\begin{eqnarray}
p\left(x,t\right) & = & \frac{1}{\sqrt{v\left(t\right)}}G\left(\frac{x-\bar{x}\left(t\right)}{\sqrt{v\left(t\right)}}\right)\\
\bar{x}\left(t\right) & = & x_{0}+\sqrt{\frac{2}{\pi}}t\\
v\left(t\right) & = & v_{0}+4\int_{0}^{t}dt^{\prime}p_{e}\left(t^{\prime}\right)\\
p_{e}\left(t\right) & = & \frac{1}{\pi}\arccos\left(q\left(t\right)\right)\\
q\left(t\right) & = & \textrm{erf}\left(\frac{\bar{x}\left(t\right)}{\sqrt{v\left(t\right)}}\right)\end{eqnarray}

Thus, the gaussian shape of the distribution is preserved, but its
center and its variance evolve in time: the center moves to the right
at constant speed, while the variance derivative is proportional to
the error rate. Convergence is thus guaranteed, since the variance
can grow at most linearly, which means that the width of the distribution
can grow at most as $\sqrt{t}$, while the center's speed is constant.
Thus, for sufficiently large times, the negative tail of the distribution,
which determines the error rate ($p_{e}\sim\sqrt{1-q}$ when $q\to1$),
will be so small that the variance will almost be constant, and this
in turn implies that the error rate decreases exponentially with time.
If we define the convergence time $T_{c}$ as the time by which the
number of wrong synapses becomes less than $1$, i.e.~when $Nq\sim1$,
we find that asymptotically $T_{c}\sim\sqrt{\log N}$, which means
that the non-rescaled convergence time is almost linear with the number
of synapses.

\begin{figure}
\includegraphics[width=0.75\columnwidth]{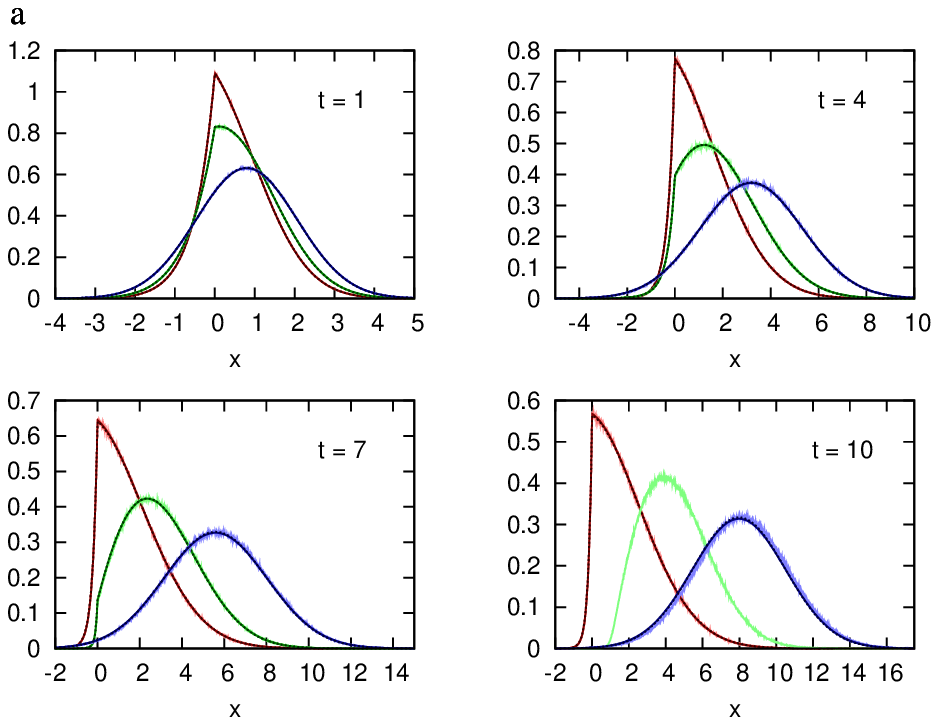} 

\includegraphics[width=0.75\columnwidth]{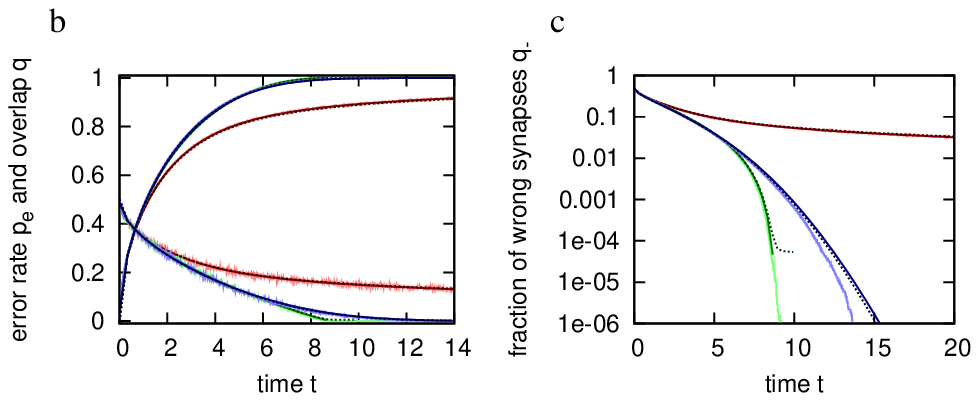} 

\caption{Comparison between simulations (light solid lines), histogram evolution
(solid lines) and continuous probability density evolution (dark dotted
lines), for three different algorithms (red: CP, green: SBPI with
$k=0.8$, blue: BPI), at different times. The curves were taken at
$N=32001$, and initialized as for Fig.~\ref{fig:hist-evol}. The
agreement between the simulations and the two analytical predictions
is almost perfect, except when $q_{-}$ is very small. \textbf{a.}
Histograms at different times. The analytical curves are not available
for SBPI at $t=10$ since at that point the algorithm has already
converged and the approximations used are no longer valid. \textbf{b.}
Average overlap $q$ (top curves, starting from 0) and error rate
$p_{e}$ (bottom curves, starting from 0.5) vs time. \textbf{c.} Fraction
of wrong synapses $q_{-}$ vs time, in logarithmic scale. This can
be used as an estimate of the convergence time with $N$; the BPI
curve is fit asymptotically by a curve which goes like $t\propto\sqrt{\log\left(q_{-}^{-1}\right)}$
(not shown).}

\label{fig:comparisons}
\end{figure}

Fig.~\ref{fig:comparisons}b shows the overlap and error rate as
a function of time; the agreement of the analytical solution with
the simulation data is almost perfect, except when $q_{-}$ is very
small, as shown in Fig.~\ref{fig:comparisons}c.

\section{Bounded hidden variables}

\label{sec:Bounded-hidden-variables} 
\begin{figure}
\includegraphics[width=0.5\columnwidth]{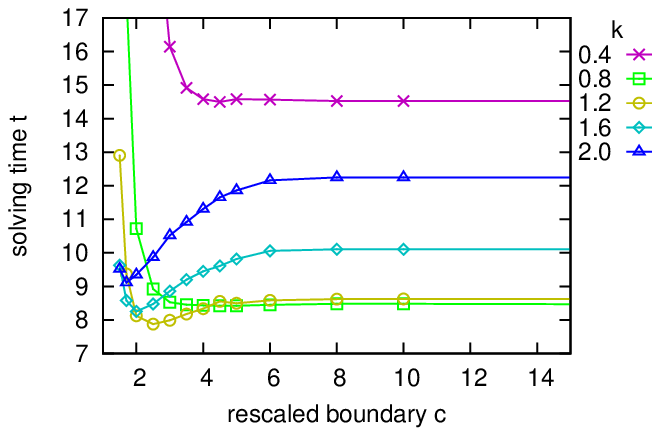} 

\caption{Solving time with different values of the rescaled boundary $c$.
Each point represents the average over 80 samples at $N=32001$ (standard
deviations are smaller than the point size). The overall optimum is
found at $c=2.5$ with $k=1.2$. These results are consistent with
those found with $N=64001$ (not shown).}

\label{fig:solving-time-with-c}
\end{figure}

We can easily introduce a limit
over the number of available hidden states, by setting a maximum value
$c\sqrt{N}$ for the modulus of $h$. Obviously, if $c$ is too small
the algorithm's performance is impaired, while if $c$ is large enough
it has no effect; in between, the behavior depends on the value of
$k$. It turns out that setting a boundary over $h$ can effectively
improve performance for BPI ($k=2$), but it has almost no effect
for the optimal SBPI algorithm, with $k\sim0.8$, similarly to what
happens in the classification problem scenario studied in \cite{baldassi07efficient}:
in fact, the optimum in this case was found to occur with $k=1.2$,
at $c=2.5$. The results are summarized in Fig.~\ref{fig:solving-time-with-c}.
An example of bounded histogram evolution is in the last panel of
Fig.~\ref{fig:hist-evol}.%

\section{Teacher with continuous synaptic weights}

\label{sec:Teacher-with-continuous-synaptic-weights} The above results
were derived in the scenario of the generalization of a learnable
rule, the desired output being provided by a binary perceptron. In
this section, we consider instead the case of a non learnable rule,
provided by a perceptron with continuous synaptic weights extracted
at random from a uniform distribution in the range $[-1,1]$: the
minimum generalization error is no longer $0$ in this case, and our
previous mean-field approach is not able to provide a simple analytic
solution; however, eq.~\ref{eq:perr} still holds true (in the limit
of large $N$), and hence the best possible assignment of the student's
weights is obtained by taking the sign of the teacher's weights, in
which case the generalization error is equal to $1/6$.

\begin{figure}
\includegraphics[width=0.75\columnwidth]{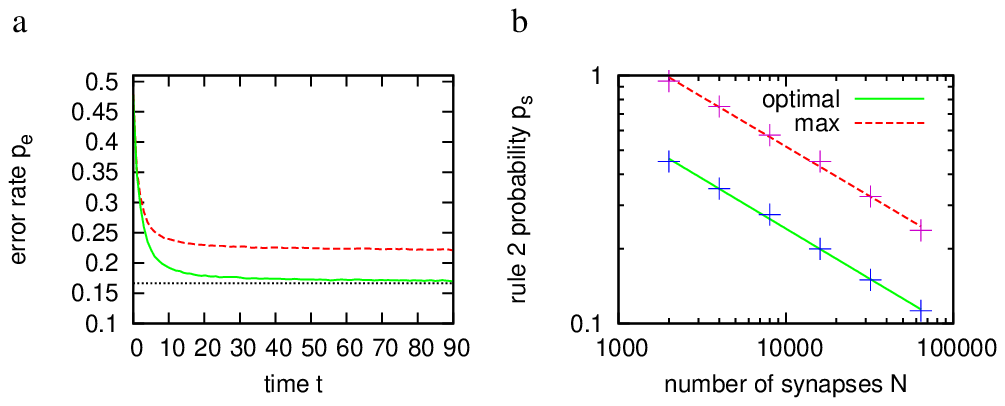}

\caption{ Simulation results using a teacher with continuous weights. \textbf{a}.
Average error rate $p_{s}$ vs time for $N=501$, $1000$ samples
(dotted black line: minimum possible error; red dashed curve: CP;
green solid curve: SBPI with optimal $p_{s}=0.8$). The learning curves
scale with $N$ following the same power law shown in the next panel.
\textbf{b}. Scaling of the $p_{s}$ parameter with $N$, in logarithmic
scale, and best fit. The fitting curves have the form $aN^{-b}$,
the fitting parameters are $a=9.9\pm0.8$, $b=0.403\pm0.008$ (optimal)
and $a=20.5\pm2.4$, $b=0.400\pm0.013$ (maximum). }

\label{fig:cont-teacher} 
\end{figure}

Our simulations show (Fig.~\ref{fig:cont-teacher}a) that even in
this case the SBPI algorithm outperforms the CP algorithm when the
parameter $p_{s}$ is chosen in the appropriate range, and that there
exists an optimal value for $p_{s}$ such that the generalization
error rapidly gets very close to the optimal value, even though the
optimum is reached in exponential time (arguably due to the fact that
the region around the solution is very flat in this case, because
some of the teacher's weights are so small that their inference is
both very difficult and not very relevant).

One important difference between this case and the previous one is
that both the optimal and the maximum value of the parameter $p_{s}$
(the maximum value is the one above which the performance becomes
equal or worse than that of CP) are not fixed with varying $N$: rather,
they both scale following the same power law (Fig.~\ref{fig:cont-teacher}b).

\section{A simplified algorithm: CP+R}

\label{sec:Simplified-algorithm} We have shown in Sec.~\ref{sec:Histogram-dynamics-for-the-SBPI-algorithm}
that, in the limit of a large number of synapses $N$ and in the context
of the generalization learning of a learnable rule, the effect of
the additional rule which distinguishes the SBPI algorithm from the
CP algorithm, and which is responsible for the superior performance
of the former with respect to the latter, is on average equivalent
to applying an unspecific, constant and low-rate meta-plastic reinforcement
to all the synapses (see eq.~\ref{eq:pb}). This reinforcement process
is only effective if it is not too strong, because otherwise it would
overcome the effect of the learning by keeping all of the synapses
away from the plastic transition boundary (i.e.~away from $x=0$,
in the notation of Sec.~\ref{sec:Continuous-limit}).

This suggests that the SBPI algorithm can be further simplified, leading
to a {}``clipped perceptron plus reinforcement'' algorithm (CP+R),
i.e.~the CP algorithm with the additional prescription that, at each
time step $\tau$, each synaptic weight undergoes a meta-plastic transition
$h_{i}^{\tau}\to h_{i}^{\tau}+2\textrm{sign}\left(h_{i}^{\tau}\right)$
with probability $p_{r}$, where $0<p_{r}<\sqrt{\frac{2}{\pi N}}$
(the time step index $\tau$ does not increment in the reinforcement
process, because it is superimposed to the standard learning rules
and acts in parallel with them). Any value of $p_{r}$ greater then
$0$ makes a qualitative difference with respect to CP.

The CP+R algorithm is only equivalent to the SBPI algorithm in the
generalization of a learnable rule scenario. Indeed, in the case of
a the non-learnable rule of Sec.~\ref{sec:Teacher-with-continuous-synaptic-weights},
the relationship of eq.~\ref{eq:pb} does not hold any more; however,
the CP+R algorithm still proves as effective as SBPI when the parameter
$p_{r}$ is properly set (not shown).

In the classification problem, on the other hand, the performance
of CP+R is worse in terms of capacity by a factor of the order of
2 with respect to SBPI. However, our preliminary results show that
the difference in such scenario between the two algorithms shows up
only in the latest phases of the learning (when the temporal correlations
in the inputs make a difference), and that simply reducing the rate
of application of the reinforcement process $p_{r}$ during the learning
along with the error rate is sufficient to recover the SBPI performance
even in that case. This will be the subject of a future work.

From the architectural point of view, such CP+R algorithm is even
simpler than the SBPI algorithm (which was already derived as a crude
simplification of the Belief Propagation algorithm); thus, it may
be an even better candidate for modelling supervised learning in biological
networks, which have very strict requirements about robustness, simplicity
and effectiveness. Its only serious drawback with respect to SBPI
is that the random reinforcement must be applied sparingly, since
the probability is of order $O\left(1/\sqrt{N}\right)$ , which would
require some fine-tuning mechanism of the cells behaviour; SBPI, on
the other hand, requires detection of near-threshold events in order
to trigger the reinforcement rule, which may also be problematic.
Furthermore, even if the learning rate under CP+R is sub-optimal with
respect to the generalization protocol problem, its extreme simplicity
and robustness might be attractive for hardware implementations of
binary perceptron units with very large number of synapses as well,
because it is adaptable to both the classification and the generalization
scenarios, and, even in the latter (algorithmically easier) case,
it greatly reduces the overhead associated with the complex computations
required by the faster algorithms, while still having a very good
scaling behaviour with $N$, as the steps required grow at most as
$\mathcal{O}\left(N\sqrt{\log N}\right)$.

\section{Summary}

\label{sec:Summary} In this paper, we have studied analytically and
through numerical simulations the SBPI algorithm dynamics in the supervised
generalization learning scenario and in the limit of a large number
of synapses $N$.

The original goal, which was that of claryfing the role of the novel
learning rule introduced by this algorithm, was approached by studying
the average dynamics of the internal synaptic state and separating
the contributions due to the different learning rules, which allowed
us to derive a partial differential equation describing the learning
process in terms of a diffusion process. The solution of such equation
in a (non-optimal) special case provided us with an estimate for the
learning time, which turned out to scale as $N\sqrt{\log N}$. The
analytical predictions were found to be in excellent agreement with
the numerical simulations.

We have also obtained some results from simulations under circumstances
in which the previous analytical approach failed, and found that the
SBPI algorithm can be further optimized by setting properly a hard
boundary to the number of internal synaptic states (scaling as $\sqrt{N}$),
which confirms our previous results in the context of classification
learning, and that its enhanced effectiveness with respect to CP is
not limited to learnable rules.

The analytical results, together with their interpretation in terms
of the synaptic states' dynamics, have also suggested the introduction
of a novel, simplified algorithm, called CP+R, which proved in our
preliminary results to be as much effective as SBPI under all the
circumstances in which we have tested it (with some minor adjustments),
making it a good candidate for biological and electronic implementations,
and which will be the subject of a future work.

\paragraph*{Acknowledgements}

I thank A. Braunstein, N. Brunel and R. Zecchina for helpful discussions
and comments on the manuscript.

\bibliographystyle{plain}
\bibliography{baldassi-generalization}

\begin{thebibliography}{10}

\bibitem{baldassi07efficient}
C.~{Baldassi}, A.~{Braunstein}, N.~{Brunel}, and R.~{Zecchina}.
\newblock Efficient supervised learning in networks with binary synapses.
\newblock {\em Proc.~Natl.~Acad.~Sci.~USA}, 104:2079--2084, 2007.

\bibitem{bhalla99}
U.~S. {Bhalla} and R.~{Iyengar}.
\newblock Emergent properties of networks of biological signaling pathways.
\newblock {\em Science}, 283:381--387, 1999.

\bibitem{bialek2000stabilitynoise}
W.~{Bialek}.
\newblock Stability and noise in biochemical switches.
\newblock {\em Adv.~Neural.~Inf.~Proc.~Sys.}, 13:103--109, 2000.

\bibitem{blum92}
A.~L. {Blum} and R.~L. {Rivest}.
\newblock Training a 3-node network is np-complete.
\newblock {\em Neural Networks}, 5:117--127, 1992.

\bibitem{braunstein06}
A.~{Braunstein} and R.~{Zecchina}.
\newblock Learning by message-passing in networks of discrete synapses.
\newblock {\em Phys. Rev. Lett.}, 96:030201, 2006.

\bibitem{engel01}
A.~{Engel} and C.~{van den Broeck}.
\newblock {\em Statistical mechanics of learning}.
\newblock Cambridge University Press, 2001.

\bibitem{fusi05}
S.~{Fusi}, P.~J. {Drew}, and L.~F. {Abbott}.
\newblock Cascade models of synaptically stored memories.
\newblock {\em Neuron}, 45(4):599--611, Feb 2005.

\bibitem{golea90}
M.~{Golea} and M.~{Marchand}.
\newblock On learning perceptrons with binary weights.
\newblock {\em Neural Computation}, 78:333--342, 1993.

\bibitem{gutfreund90b}
H.~{Gutfreund} and Y.~{Stein}.
\newblock Capacity of neural networks with discrete synaptic couplings.
\newblock {\em J.~Phys.~A: Math.~Gen.}, 23:2613--2630, 1990.

\bibitem{hayer05molecularswitches}
A.~{Hayer} and U.~S. {Bhalla}.
\newblock Molecular switches at the synapse emerge from receptor and kinase
  traffic.
\newblock {\em PLoS Comput.~Biol.}, 1(2):e20, 2005.

\bibitem{kabashima04}
Y.~{Kabashima} and S.~{Uda}.
\newblock {\em A BP-based algorithm for performing bayesian inference in large
  perceptron-type networks}, volume 3244, pages 479--493.
\newblock Springer Berlin / Heidelberg, 2004.

\bibitem{kinouchi92optimalgeneralization}
O.~{Kinouchi} and N.~{Caticha}.
\newblock Optimal generalization in perceptrons.
\newblock {\em J.~Phys.~A}, 25:6243, 1992.

\bibitem{krauth89}
W.~{Krauth} and M.~{M{\'e}zard}.
\newblock Storage capacity of memory networks with binary couplings.
\newblock {\em J.~Phys.~France}, 50:3057, 1989.

\bibitem{miller05}
P.~{Miller}, A.~M. {Zhabotinsky}, J.~E. {Lisman}, and X.-J. {Wang}.
\newblock The stability of a stochastic camkii switch: dependence on the number
  of enzyme molecules and protein turnover.
\newblock {\em PLoS Biol}, 3(4):e107, Mar 2005.

\bibitem{oconnor05}
D.~H. {O'Connor}, G.~M. {Wittenberg}, and S.~S.-H. {Wang}.
\newblock Graded bidirectional synaptic plasticity is composed of switch-like
  unitary events.
\newblock {\em Proc. Natl. Acad. Sci. USA}, 102:9679--9684, 2005.

\bibitem{petersen98}
C.~C. {Petersen}, R.~C. {Malenka}, R.~A. {Nicoll}, and J.~J. {Hopfield}.
\newblock All-or-none potentiation at {CA3-CA1} synapses.
\newblock {\em Proc. Natl. Acad. Sci. USA}, 95:4732--4737, 1998.

\bibitem{rosenzvi00}
M.~{Rosen-Zvi}.
\newblock On-line learning in the ising perceptron.
\newblock {\em J.~Phys.~A}, 33:7277--7287, 2000.

\bibitem{rosenblatt62}
F.~{Rosenblatt}.
\newblock {\em Principles of neurodynamics}.
\newblock Spartan Books, New York, 1962.

\bibitem{Solla98optimalperceptron}
S.~A. {Solla} and O.~{Winther}.
\newblock Optimal perceptron learning: an online bayesian approach.
\newblock In {\em On-Line Learning in Neural Networks. Combridge}. University
  Press, 1998.

\bibitem{yedidia03}
J.~S. {Yedidia}, W.~T. {Freeman}, and Y.~{Weiss}.
\newblock {\em Understanding Belief Propagation and its generalizations},
  chapter~8, pages 236--239.
\newblock Morgan Kaufman, 2003.

\end{thebibliography}

\end{document}